\documentclass[twocolumn,showpacs,preprintnumbers,amsmath,amssymb,prb,superscriptaddress]{revtex4-1}

\usepackage{amssymb}
\usepackage{graphicx}
\usepackage{bm}

\begin{document}
\title{Antiferromagnetic critical pressure in URu$_2$Si$_2$ under hydrostatic conditions}

\author{Nicholas P. Butch}
\affiliation{Center for Nanophysics and Advanced Materials, Department of Physics, University of Maryland, College Park, MD 20742}
\email{nbutch@umd.edu}
\author{Jason R. Jeffries}
\affiliation{Condensed Matter and Materials Division, Lawrence Livermore National Laboratory, Livermore, CA 94550}
\author{Songxue Chi}
\author{Juscelino Batista Le\~{a}o}
\author{Jeffrey W. Lynn}
\affiliation{NIST Center for Neutron Research, National Institute of Standards and Technology, Gaithersburg, MD 20899}
\author{M. Brian Maple}
\affiliation{Department of Physics, University of California, San Diego, La Jolla, CA 92093}
\date{\today}

\begin{abstract}
The onset of antiferromagnetic order in URu$_2$Si$_2$ has been studied via neutron diffraction in a helium pressure medium, which most closely approximates hydrostatic conditions. The antiferromagnetic critical pressure is 0.80~GPa, considerably higher than values previously reported. Complementary electrical resistivity measurements imply that the hidden order-antiferromagnetic bicritical point falls between 1.3 and 1.5~GPa. Moreover, the redefined pressure-temperature phase diagram suggests that the superconducting and antiferromagnetic phase boundaries actually meet at a common critical pressure at zero temperature.

\end{abstract}

\maketitle

For over two decades, the heavy fermion superconductor URu$_2$Si$_2$ has challenged researchers. The hidden order (HO) state with transition temperature $T_0 = 17.5$~K is characterized by a BCS-like specific heat anomaly\cite{Maple86} that is too large to be due solely to a small associated antiferromagnetic (AFM) moment,\cite{Broholm87} which is now believed to arise from internal strain due to sample defects. Many bulk property measurements evince that the HO transition leads to a partial gapping of the Fermi surface: the BCS-like specific heat anomaly, as well as anomalies in the electrical resistivity, magnetic susceptibility,\cite{Maple86,Palstra85,Schlabitz86} ultrasound,\cite{Kuwahara97} thermal expansion,\cite{Visser86} and lattice thermal conductivity.\cite{Behnia05,Sharma06} Gaps are also observed in tunneling\cite{Hasselbach92} and spin excitation spectra.\cite{Wiebe07} Yet, the identity of the microscopic order parameter remains a contentious, open question.

Although the HO phase is thought to be nonmagnetic, it is fairly unstable against magnetic order, which can be induced by light chemical substitution,\cite{Miyako92,Bauer05,Butch09b,Butch10} and AFM order requires relatively little applied pressure $P$. While a kink in $T_0(P)$ at 1.5~GPa was seen in early studies,\cite{McElfresh87} intrinsic long range AFM order was first identified above a similar critical pressure $P_c$ via neutron diffraction.\cite{Amitsuka99} Soon thereafter, NMR and $\mu$SR studies established that HO and AFM phases inhomogeneously coexist at much lower $P$.\cite{Matsuda01,Amato04} Subsequent neutron diffraction reports identified $P_c$ in the range of $0.4-0.7$~GPa,\cite{Bourdarot05,Amitsuka07,Amitsuka08,Niklowitz10} which was corroborated by specific heat and transport measurements.\cite{Hassinger08,Motoyama08} These studies also showed that the HO-AFM phase boundary meets the $T_0$ line at a bicritical point, implying that the HO and AFM order parameters are uncoupled, and that the two phases must be separated by a first-order phase transition. In addition, magnetic susceptibility\cite{Amitsuka07,Uemura05} and specific heat\cite{Hassinger08} measurements suggested that bulk superconductivity ($T_c=1.3$~K) is suppressed discontinuously at $P_c$, where AFM order arises, and that resistivity studies at higher $P$ only probed filaments or patches.\cite{Hassinger08,Jeffries07,Jeffries08}

The apparent current consensus is that $P_c \approx 0.5$~GPa,\cite{Amitsuka08,Niklowitz10,Hassinger08,Motoyama08} but there is good reason to accept this value cautiously. Several studies have shown the importance of experimental conditions during pressure measurements on URu$_2$Si$_2$. The most pronounced example is that when measured in He, the most hydrostatic pressure medium available, no AFM moment appeared below 0.5~GPa, but in the same sample a substantial AFM moment was detected at 0.45~GPa when using the less hydrostatic Fluorinert liquid.\cite{Bourdarot05} A variation in $P_c$ was also shown between Fluorinert and Daphne oil,\cite{Amitsuka08} and clear pressure hysteresis in the value of $T_0$ was demonstrated in Fluorinert.\cite{Butch09} The sensitivity of URu$_2$Si$_2$ to nonhydrostatic conditions is further exemplified by its dramatically anisotropic response to uniaxial stress.\cite{Yokoyama05} It is worth noting that URu$_2$Si$_2$ is not unique in this regard; it was recently demonstrated in the structurally related iron pnictide compound CaFe$_2$As$_2$ that nonhydrostatic conditions can result in phase coexistence.\cite{Goldman09} Motivated by these earlier studies, we investigated the onset of AFM order in URu$_2$Si$_2$ using a He cell with a maximum working pressure of 1.0~GPa, doubling the range measured in Ref.~\onlinecite{Bourdarot05}. We find that hydrostatic conditions are of paramount importance. The AFM transition is sharper and $P_c = 0.80$~GPa, which provocatively matches the superconducting (SC) critical endpoint inferred from earlier measurements\cite{Jeffries07,Jeffries08} and intimates the existence of a zero temperature multicritical point at $P_c$.

A single crystal of URu$_2$Si$_2$ was grown via the Czochralski technique in an electric tetra-arc furnace and annealed in Ar for one week. Samples were oriented via x-ray Laue back reflection or monochromatic x-ray diffraction, and cut with a low-speed diamond wheel saw. Transmission electron microscopy confirmed crystallographic purity, with extremely low dislocation density $10^1$~mm$^{-2}$, while energy dispersive spectroscopy showed no presence of extra elements; a sample was further characterized via electron energy loss spectroscopy.\cite{Jeffries10} Neutron diffraction measurements were performed on a 1~g single crystal at the NIST Center for Neutron Research on the BT-9 triple axis spectrometer with 14.7 meV incident energy ($\lambda = 2.36$~\AA), a pyrolytic graphite filter, and 40'-48'-40'-open collimation. Temperature was controlled by a He cryostat and pressure was applied using a He-gas cell connected to a two-stage intensifier through a heated capillary. Pressure was adjusted only at temperatures well above the He melting curve, and the capillary was heated during slow cooling of the cell to accommodate the contracting He, minimizing pressure loss. Measurements of electrical resistivity $\rho$, with current in the basal plane, were performed in a piston-cylinder cell in a commercial cryostat using a 1:1 volume mixture of n-pentane / isoamyl alcohol. The superconducting transition of Sn was used as a manometer.

\begin{figure}
    {\includegraphics[width=3.25in]{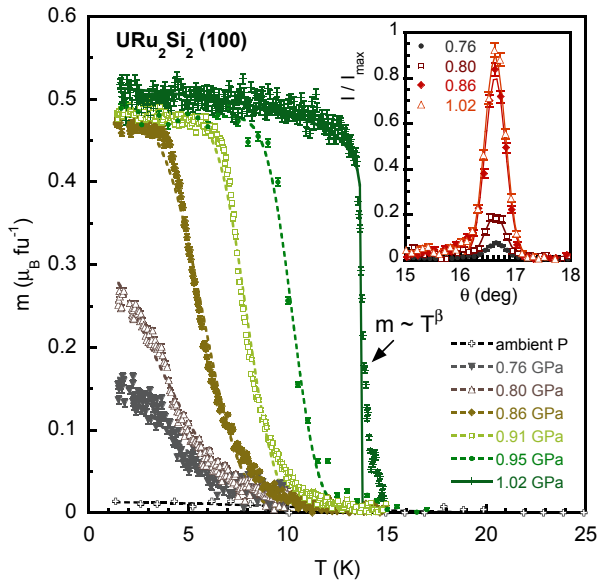}}
    \caption{(Color online) Temperature dependence of the AFM moment under hydrostatic pressure. A fit (solid line) of the form $m \sim T^\beta$ to the 1.02~GPa data yields a minuscule $\beta = 0.05$. The sharp onset and weak slope in the ordered phase are suggestive of a discontinuous transition. Error bars represent one standard deviation. Dotted lines are guides to the eye. Inset: Rocking scans at 1.5~K normalized to the intensity at 1.02~GPa.}
    \label{moments}
\end{figure}

To characterize the onset of AFM order, the intensity of the magnetic (100) peak, a forbidden nuclear reflection, was compared to the intensity of the nuclear (200) peak, which was constant between 30 and 1.5~K. The magnitude of the ordered moment was calibrated at 81.8 meV incident energy, where extinction of the strong (200) peak was negligible. Rocking scans of the (100) peak at 1.5~K are shown in the inset of Fig.~\ref{moments}, with intensities normalized to the value at the highest pressure of 1.02~GPa. The temperature dependence of the ordered moment $m(T)$ is shown in Fig.~\ref{moments}. At low $T$ and ambient pressure $m=0.011$~$\mu_B$ per formula unit, comparable to recently reported values,\cite{Amitsuka07,Niklowitz10} and grows by a factor of almost 50 by 1.02~GPa. A fit of the form $m \sim T^\beta$ to the 1.02~GPa data yields an exponent $\beta = 0.05$ that is too small to describe a conventional continuous transition and is consistent with a first order transition. Compared to previous reports,\cite{Amitsuka99,Bourdarot05,Amitsuka07,Niklowitz10} these $m(T)$ curves exhibit less curvature in the ordered state and narrower transitions in $T$. This sharper discontinuity can be attributed to more ideal hydrostatic conditions in the He cell, resulting in less smearing of the transition. No hysteresis in $T$ was observed. The moment $m=0.52$~$\mu_B$ determined here is modestly larger than the previously reported value of 0.4~$\mu_B$.\cite{Amitsuka07} However, a 0.52~$\mu_B$ static moment is still small compared to the 1.2~$\mu_B$ transition moment of the (100) spin excitation at ambient pressure.\cite{Broholm91}

\begin{figure}
    {\includegraphics[width=3.25in]{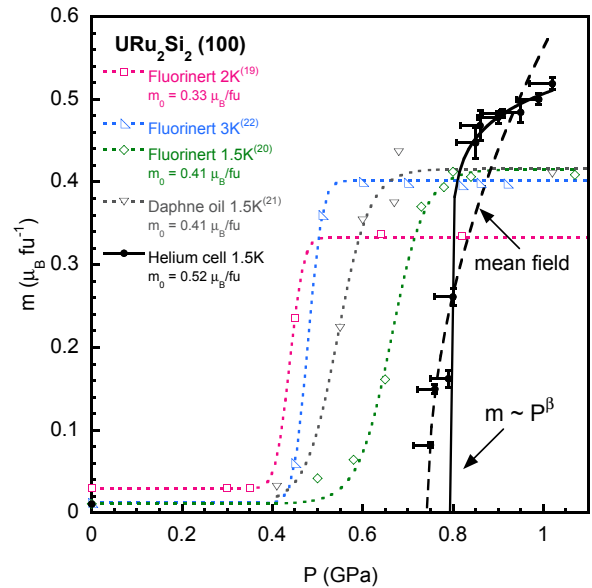}}
    \caption{(Color online) Low temperature onset of AFM order. In He, the onset occurs at significantly higher $P$ than in less hydrostatic media. The solid line is a fit of the form $m \sim P^\beta$ yielding $\beta = 0.08$. The dashed line is a mean field fit using fixed $\beta = 0.5$. Error bars in $m$ represent one standard deviation while error bars in $P$ represent a 5\% uncertainty. Dotted lines are guides to the eye. Ref.~\onlinecite{Niklowitz10} assumes $m=0.4$~$\mu_B$.}
    \label{mP}
\end{figure}

The increase in the ordered moment at $T = 1.5$~K is shown in Fig.~\ref{mP}. Vertical error bars represent one standard deviation, while the horizontal error bars reflect a 5\% uncertainty in $P$ due to contraction of He at low $T$. These data illustrate the sudden zero-$T$ onset of AFM order from the HO state. Between 0.75 and 0.85~GPa, the slope $\frac{\partial{m}}{\partial{P}} = 3.5$~$\mu_B$~fu$^{-1}$~GPa$^{-1}$ before it starts to saturate above 0.85~GPa. A fit of the form $m \sim P^\beta$ yields $\beta = 0.08$, which is similar to the $m(T)$ exponent and points to a discontinuous AFM onset at low $T$. A mean-field fit poorly describes the data and is clearly inapplicable. The midpoint of the transition is used to define the zero-temperature critical pressure $P_c= 0.80(1)$~GPa. Similar $m(P)$ data from several recent neutron diffraction studies are presented for comparison, showing that in He, the value of $P_c$ is the highest by a significant margin. The discrepancies in $P_c$ between data sets are attributable to the less hydrostatic media used: Fluorinert\cite{Bourdarot05,Niklowitz10,Amitsuka07} and Daphne oil.\cite{Amitsuka08}

\begin{figure}
    {\includegraphics[width=3.25in]{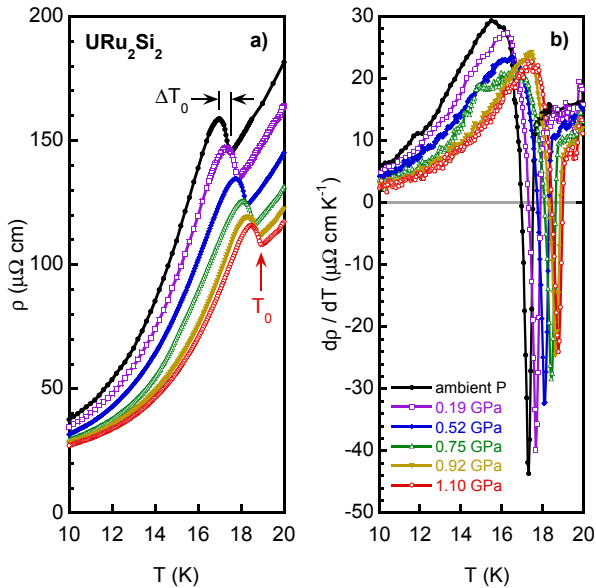}}
    \caption{(Color online) Pressure dependence of electrical resistivity. a) The HO transition, marked by a sharp anomaly at $T_0$, increases with $P$.  b) Temperature derivative of $\rho(T)$, highlighting the narrow transition width and the lack of any obvious features associated with the AFM transition at $T_x$.}
    \label{rho}
\end{figure}

\begin{figure}
    {\includegraphics[width=3.25in]{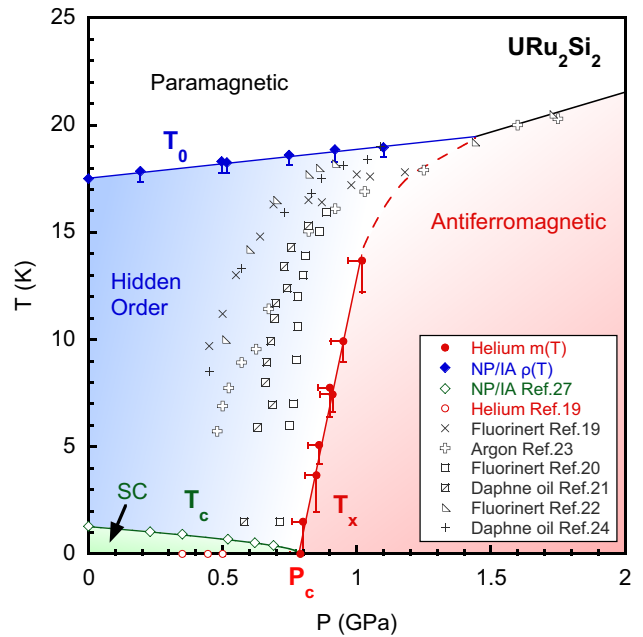}}
    \caption{(Color online) Pressure-temperature phase diagram of URu$_2$Si$_2$ in helium. $T_0$ is defined by the local minimum in the $\rho(T)$ data; the error bar represents $\Delta T_0$. $T_x$ is defined where $m(T)$ reaches half of its full value, the vertical error bar 90\%. The horizontal error bar represents a 5\% uncertainty in $P$. A comparison to published data shows that the value of the AFM critical pressure $P_c$ is substantially higher under hydrostatic conditions.\cite{Bourdarot05,Hassinger08,Motoyama08,Amitsuka07,Amitsuka08,Niklowitz10} The redefined $T_x$ suggests that SC and AFM phases meet at 0~K. }
    \label{phsdgm}
\end{figure}

The paramagnetic(PM)-HO transition temperature $T_0$ was determined via electrical resistivity measurements on a small piece of the URu$_2$Si$_2$ crystal. The $P$ dependence of the $\rho(T)$ data is shown in Fig.~\ref{rho}. The anomaly takes the form of a peak-trough structure with a sharp local minimum, by which $T_0$ is defined; the magnitude of $\frac{\partial{\rho}}{\partial{T}}$ is exhibited in Fig.~\ref{rho}b. Applied pressure enhances $T_0$, which has a linear $P$ dependence, and reduces $\rho(T_0)$, although the width $\Delta T_0$ of the transition, from minimum to maximum, decreases only slightly. These properties are all consistent with previous studies. Below 10~K, $\rho(T)$ is best described by a power law with exponent approximately 1.7. In some recent studies, $\rho(T)$ data have also shown anomalies at the pressure induced HO-AFM transition at $T_x$, although their magnitudes, or even detectable presence, are sample dependent.\cite{Motoyama08} These anomalies are changes in slope most easily identified as secondary peaks in $\frac{\partial{\rho}}{\partial{T}}$ at temperatures less than $T_0$.\cite{Hassinger08,Motoyama08} In the present study, the systematic evolution of such features is absent in two important $P$ ranges: at 0.92 and 1.10~GPa, where they would be expected based on the neutron diffraction data (c.f. Fig.~\ref{phsdgm}), and from 0.52 to 1.10~GPa, where they would be expected if $T_x$ matched previous studies.\cite{Hassinger08,Motoyama08,Amitsuka08,Niklowitz10} At 0.75~GPa, an extra anomaly in $\frac{\partial{\rho}}{\partial{T}}$ at 15~K is attributable to the loss of hydrostaticity in a leaking cell just prior to failure. If this anomaly really reflected the onset of AFM order, it should also appear at higher $P$, but it is conspicuously absent.

A $P-T$ phase diagram based on the $m(T)$ and $\rho(T)$ data is presented in Fig.~\ref{phsdgm}. Following Refs.~\onlinecite{Amitsuka07,Amitsuka08}, the HO-AFM phase boundary $T_x$ is defined where 50\% of the full moment is observed, while the error bars indicate where $m(T)$ is 90\% of its full value. Due to the high $T$ tails in $m(T)$ (Fig.~\ref{moments}) a 10\% criterion is not directly associable with the onset of long range order and is not shown. This phase boundary has a linear slope $\frac{\partial{T_x}}{\partial{P}}=58$~K~GPa$^{-1}$, extrapolating to 0.78(5)~GPa. The $T_0$ boundary, with error bars indicating $\Delta T_0$, is also linear with a slope $\frac{\partial{T_0}}{\partial{P}}=1.3$~K~GPa$^{-1}$. The $T_0$ boundary seems robust between different reports, and in fact our $T_0$ line agrees very well with that of Ref.~\onlinecite{Hassinger08}. The $T_0$ and $T_x$ lines extrapolate to an intersection at (1.1~GPa, 19~K), but based on most previous reports, the $T_x$ boundary actually curves and meets $T_0$ at about 1.5~GPa. This could not be directly confirmed in the present study because of the 1~GPa limit of the pressure cell.

A comparison of the $T_x$ boundary determined from our data to previous reports shows clearly that it occurs at significantly higher pressure in He. As shown in Fig.~\ref{phsdgm}, the other reported transitions occur between 0.5 and 0.7~GPa, with the obvious exception of the data taken in He by Bourdarot and coworkers, which show no moment up to 0.5~GPa.\cite{Bourdarot05} These studies represent a variety of probes of the structural\cite{Hassinger08,Motoyama08,Niklowitz10} and magnetic\cite{Bourdarot05,Amitsuka07,Amitsuka08,Niklowitz10} lattices. It is important to note that the dissimilitude between the various reported phase boundaries is not primarily due to sample dependence or measurement technique. The HO-AFM transitions determined via thermal expansion and Larmor diffraction have been shown to match those defined using neutron diffraction data.\cite{Villaume08,Niklowitz10} As already noted, it has also been demonstrated that for the same sample, the choice of pressure medium causes a pronounced variation in the AFM onset.\cite{Bourdarot05,Amitsuka08} The higher value of $P_c$ determined in our study is thus attributable inherently to better hydrostatic conditions.

There are several implications of the redefined phase boundary. Although it has already been established that the HO-AFM and PM-HO boundaries meet at a multicritical point, it has been identified at three different pressures: 0.9~GPa,\cite{Niklowitz10} 1.09~GPa,\cite{Motoyama08} and 1.3~GPa.\cite{Hassinger08} From our measurements in He, it is clear that these boundaries actually meet at $P > 1.02$~GPa. Moreover, given the known reduction in $\frac{\partial{T_x}}{\partial{P}}$ at higher $P$, an intersection between 1.3 and 1.5~GPa is most likely. The higher value of $P_c$ also implies that bulk SC meets the HO-AFM boundary at 0~K. This is illustrated in Fig.~\ref{phsdgm} using the SC $10\% \rho(T)$ transition from Ref.~\onlinecite{Jeffries08} as an indicator of bulk SC, which tracks  well data at lower $P$ from bulk probes: specific heat\cite{Hassinger08} and magnetic susceptibility.\cite{Amitsuka08} The main difference in a hydrostatic environment is that $T_c$ is suppressed continuously to 0~K and does not intersect the HO-AFM boundary at finite $T$, as it does in less hydrostatic pressure media.\cite{Hassinger08,Amitsuka08} Thus, the reported $P$-driven discontinuous SC phase transition\cite{Hassinger08,Amitsuka08} is not intrinsic, but likely due to the premature onset of AFM order arising from a nonhydrostatic environment.

Most tantalizingly, the endpoint of the SC phase boundary extrapolates to $P_c$, which suggests that the SC pairing energy scale goes to zero exactly at the onset of long-range AFM. The abruptness of the low-$T$ onset of $m(P)$ appears to exclude a scenario where SC arises due to AFM critical fluctuations. However, it is tempting to speculate that the (100) AFM magnetic fluctuation spectrum, which has been shown to disappear at the onset of AFM,\cite{Villaume08} is related to the SC state. That the gap remains finite when the intensity of the excitations disappears at 0.67~GPa may be due to the early onset of AFM order arising from nonhydrostatic conditions. Moreover, the inherent anisotropy of these magnetic excitations would conveniently explain the large anisotropy in the SC upper critical field, which is significantly smaller along the magnetic easy tetragonal $c$ axis.\cite{Jeffries07} Under pressure, the intensity of these fluctuations should diminish, tracking the suppression of $T_c$, despite the increase of $T_0$. The pressure dependence of these excitations has not yet been studied in detail, but could offer important clues to the relationship between the HO and SC phases.

To summarize, the onset of AFM order in URu$_2$Si$_2$ was studied in a He pressure cell, yielding $P_c = 0.80$~GPa, a value significantly higher than previously reported, which underscores the importance of hydrostatic measurement conditions. Electrical resistivity measurements, while insensitive to the $T_x$ transition, indicate that the PM-HO-AFM bicritical point lies between 1.3 and 1.5~GPa. The redrawn $P-T$ phase diagram shows that $T_c$ is suppressed to zero at $P_c$, suggesting a more intimate relationship between SC and AFM phases.

\begin{acknowledgments}
Sample preparation was supported by the DOE under Research Grant \# DE-FG02-04ER46105. LLNL is operated by Lawrence Livermore National Security, LLC, for the DOE, NNSA under Contract DE-AC52-07NA27344. J.R.J. is supported by the Science Campaign at LLNL. N.P.B. is supported by CNAM.
\end{acknowledgments}

\end{document}